\documentclass[aip,rsi,amsmath,amssymb,reprint]{revtex4-1}

\usepackage{graphicx}
\usepackage{dcolumn}
\usepackage{bm}
\usepackage{color} 

\definecolor{orange2}{RGB}{248,152,37}
\definecolor{brown2}{RGB}{190,125,42}
\definecolor{purple2}{RGB}{171,74,156}
\definecolor{green2}{RGB}{0,166,81}
\definecolor{blue2}{RGB}{58,60,151}
\definecolor{yellow2}{RGB}{255,242,0}
\definecolor{red2}{RGB}{237,28,36}
\definecolor{black2}{RGB}{35,31,32}

\begin{document}

\title[Integrated Capacitance Bridge]{An integrated capacitance bridge for high-resolution, wide temperature range quantum capacitance measurements}

\author{Arash Hazeghi}
\affiliation{These authors contributed equally to this work}
 \affiliation{Department of Electrical Engineering, Stanford University}

\author{Joseph A. Sulpizio}
\affiliation{These authors contributed equally to this work}
\affiliation{Department of Physics, Stanford University}

\author{Georgi Diankov}
\affiliation{Department of Chemistry, Stanford University}

\author{David Goldhaber-Gordon}
\email{goldhaber-gordon@stanford.edu}
\affiliation{Department of Physics, Stanford University}

\author{H.S. Philip Wong}
 \affiliation{Department of Electrical Engineering, Stanford University}
\date{\today}

\begin{abstract}
We have developed a highly-sensitive integrated capacitance bridge for quantum capacitance measurements.  Our bridge, based on a GaAs HEMT amplifier, delivers attofarad (aF) resolution using a small AC excitation at or below $k_{B}T$ over a broad temperature range (4K-300K).  We have achieved a resolution at room temperature of $\mathrm{10 aF}/\sqrt{\mathrm{Hz}}$ for a $\mathrm{10 mV}$ AC excitation at 17.5 kHz, with improved resolution at cryogenic temperatures, for the same excitation amplitude.  We demonstrate the performance of our capacitance bridge by measuring the quantum capacitance of top-gated graphene devices and comparing against results obtained with the highest resolution commercially-available capacitance measurement bridge.  Under identical test conditions, our bridge exceeds the resolution of the commercial tool by up to several orders of magnitude.
\end{abstract}

\maketitle

\section{Introduction}

As device scaling continues below 20 nm and novel nanodevices appear on the horizon, accurate characterization and detailed understanding of their electronic structure is essential, yet challenging\cite{itrs2009}.  For example, carrier transport through a nanostructure, which is often quantified by mobility, cannot be accurately characterized from conductance alone without knowing the exact carrier density.  Information on carrier density is often extracted from a capacitance spectrum by measuring the capacitance between the device channel and a gating terminal as a function of DC bias, commonly known as a CV curve.

In most field-effect semiconductor devices, a dielectric layer isolates the channel from the gate electrode.  The capacitance measured from the gate is the series combination of the geometric capacitance associated with the dielectric, $C_{ox}$, and the capacitance associated with adding carriers to the bandstructure of the semiconductor, or the quantum capacitance, $C_Q$, which is proportional to the electronic density of states (DOS)\cite{datta2007,Lee1982,luryi}.  For devices with large DOS in the channel, the effective gate capacitance is simply the geometric capacitance $C_{ox}$.  However, in nanoscale devices with strongly coupled gates, a low DOS in the channel can reduce the quantum capacitance to hundreds of attofarads, making $C_Q$ dominate the total gate capacitance.  In this regime, the total capacitance is a strong function of the channel DOS.  In order to fully resolve fine bandstructure features, for example van Hove singularities in carbon nanotubes\cite{Ilani2006}, an external excitation smaller than the characteristic thermal energy $k_{B}T$ is necessary.  Inevitably, any practical measurement setup will include some length of cables that have a finite parasitic capacitance on the order of hundreds of picofarads.  This produces an enormous attenuation of the test signal, pushing even state-of-the-art laboratory CV meters beyond their resolution limits when such small test signals are used, as illustrated in Fig.~\ref{fig:schema}a.

In this work, we present an integrated capacitance bridge to extract and balance the test signal coming from a nanostructure before it is attenuated by external cables.  We demonstrate excellent capacitance resolution for test signals less than $k_{B}T$ from room temperature down to temperatures of 4K, yielding an output noise of less than $10 \mathrm{nV}/\sqrt{\mathrm{Hz}}$.  We compare our device to a commercially available, state-of-the-art capacitance/loss measurement tool, and are able to significantly exceed its performance when measuring the capacitance of top-gated graphene devices.

\section{Bridge Design and Operation}

\begin{figure*}[htpb]
\includegraphics[width=17cm]{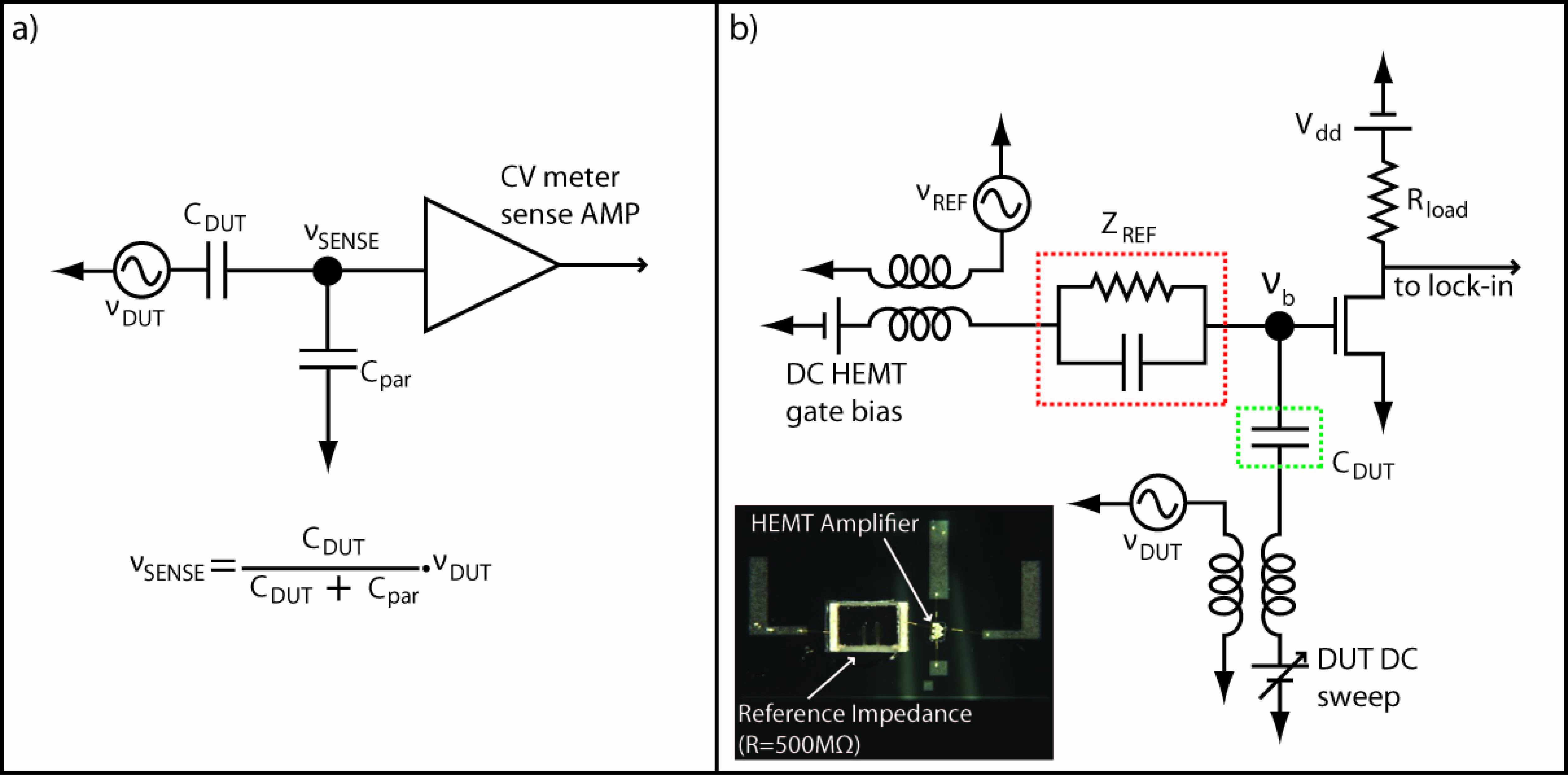}
\caption{\label{fig:schema} a) Parasitic cable capacitance in a typical capacitance test setup with a CV meter. For small $C_{DUT}$, the test signal $\nu_{DUT}$ is attenuated by a factor of $\sim C_{DUT}/C_{par}$, which is on the order of 10$^6$ for typical nanostructures connected via standard coaxial cables.  For $\nu_{DUT}\sim k_{B}T$ at 4K, this results in $\nu_{sense}<$ 1nV.  b) Schematics of our bridge circuit connected to a DUT.  AC and DC signals are added together with a Triad Magnetics SP-67 audio transformer.  The inset contains a photograph of the bridge on the GaAs substrate before wirebonding to the DUT.}
\end{figure*}

The principles of operation of the capacitance bridge are explained by Steele \textit{et al.}\cite{steele}  The bridge consists of a reference impedance and an impedance-matching amplifier, whose function is to drive the large parasitic cable capacitance and isolate the device under test (DUT).  Since we are interested in measurements across a broad temperature range down to cryogenic temperatures, we use a GaAs-based high electron mobility transistor (HEMT) as the impedance-matching amplifier\cite{urazhdin2002}.  A standard Si FET is unsuitable due to carrier freeze-out at low temperature.  The HEMT in our bridge is an unpackaged FHX35X transistor manufactured by the Fujitsu Corp, and has a wide ($\sim\!280 \mathrm{\mu m}$) channel fabricated from epitaxially-grown GaAs, with a gate capacitance $\sim\!0.4 \mathrm{pF}$\cite{resnote}. The 2D electron gas is fully depleted when the gate is biased at -1V (depletion mode).

The reference impedance is used to balance the signal across the DUT, and its AC impedance must be larger than the HEMT gate AC impedance to avoid shunting of the DUT signal.  As the HEMT gate is DC biased through the reference (via a Yokogawa 7651 programmable low-noise DC power supply), the reference DC impedance should ideally be at most of the same order as the HEMT gate resistance across all temperatures.  To satisfy these constraints at the test frequency 17.5 kHz, we use a 500 M$\Omega$ thick-film resistor (Tyco Electronics part no. 26M2248) with a low thermal coefficient and a parasitic capacitance $C_{REF}\sim\! 110 \mathrm{fF}$. A 1k$\Omega$ load resistor (Vishay Dale no. CCF551K00FKE36) is used to bias the HEMT drain.  Bridge schematics and a photograph are shown in Fig.~\ref{fig:schema}b.

To avoid strain and thermal gradients at low temperature, a thermally-matched substrate is required to mount the bridge circuit.  We used a semi-insulating GaAs wafer as the substrate, onto which a 23nm Al$_{2}$O$_{3}$ layer was grown via ALD for additional electrical isolation.  Standard photolithography/liftoff processing were used to fabricate 300nm thick Al electrodes for bonding.  The bridge components were then attached to the substrate, using thermally-conductive silver epoxy for the HEMT and PMMA for the reference resistor, then wire-bonded to the pads.

\begin{figure}
\includegraphics[width=8.5cm]{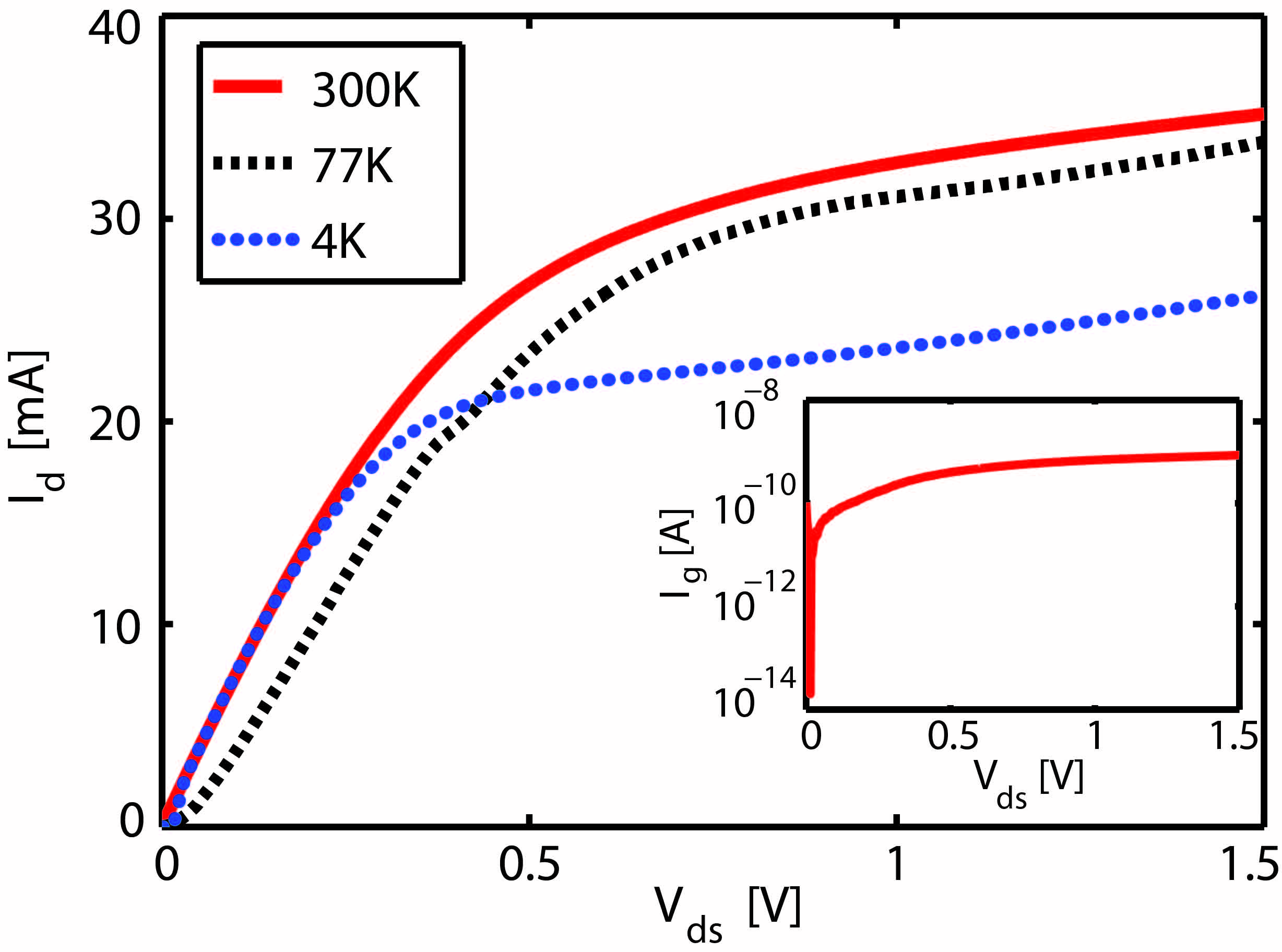}
\caption{\label{fig:hemtfig} Output characteristics of Fujitsu FHX35X HEMT for $V_{gs}=0$ at various temperatures.  The inset shows the gate current $I_{g}$ as a function of $V_{ds}$ for $V_{gs}=0$ (no load resistor).  For temperatures 77K and below, the leakage is below 1pA.}
\end{figure}

\begin{figure}
\includegraphics[width=8.5cm]{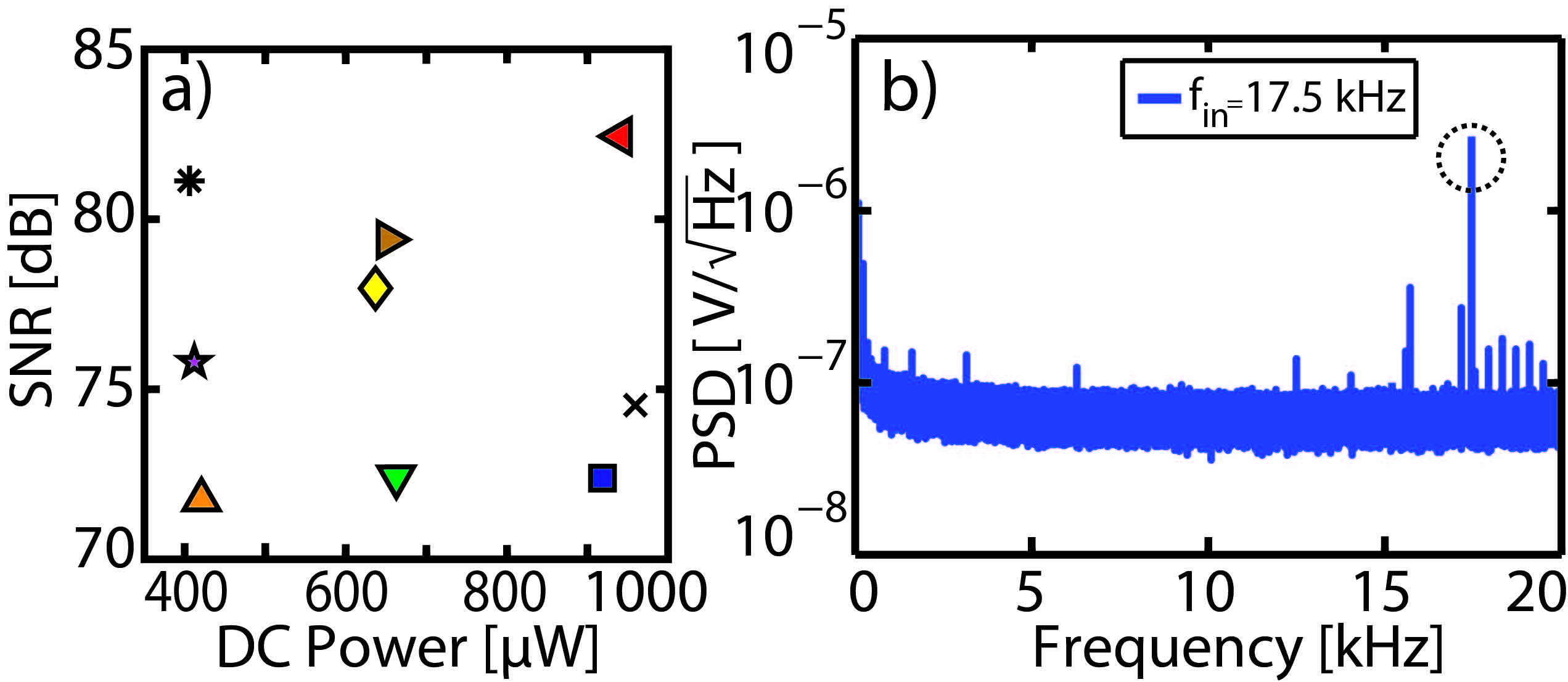}
\caption{\label{fig:noisefig} a) Integrated bridge output SNR as a function of DC power dissipation for a 1.5mV RMS input signal at room temperature.  The symbols in the plot correspond to the following $(V_{gs}[V],V_{dd}[V])$ pairs: \textcolor{black2}{$\ast$} (-0.05,4), \textcolor{purple2}{$\bigstar$} (-0.1,4), \textcolor{orange2}{$\blacktriangle$} (-0.15,4), \textcolor{brown2}{$\blacktriangleright$} (-0.1,5), \textcolor{yellow2}{$\blacklozenge$} (-0.05,5), \textcolor{green2}{$\blacktriangledown$} (-0.15,5), \textcolor{red2}{$\blacktriangleleft$} (-0.1,6), $\times$ (-0.15,6), \textcolor{blue2}{$\blacksquare$} (-0.05,6).  The optimal bias point, i.e. high SNR and low power, is marked by $\ast$.  b) output power spectral density (PSD) of the bridge (measured with a home-built spectrum analyzer) biased at the optimal bias point(-60 dB gain between bridge input and output) with 100$\mu$V RMS input signal).  The excitation at 17.5kHz is indicated by the black circle. }
\end{figure}
\begin{table*}
\caption{\label{tab:tab1}Optimized HEMT bias conditions with corresponding amplifier output sensitivity $S$ and best achievable capacitance resolution $\delta \tilde{C}$ (measured using $C_{DUT}$=250aF, before bonding graphene device to bridge).}
\begin{ruledtabular}
\begin{tabular}{cccccc}
 $T$[K]&$V_{dd}$[V]&$V_{gs}$[mV]&$S$[nV/$\sqrt{\mathrm{Hz}}$]&HEMT Gain&$\delta \tilde{C}$ ($\nu_{DUT}=$10mV rms) [aF/$\sqrt{\mathrm{Hz}}$]\\
\hline
300& 4 & -50 & 8 & 0.05 & 9.6 \\
77 & 4 & -50 & 7.4 & 0.063 & 7\\
4.2 & 4 & -25 & 5.6 & 0.1 & 3.4\\
\end{tabular}
\end{ruledtabular}
\end{table*}
In operation, two out-of-phase AC signals, $\nu_{DUT}$ and $\nu_{REF}$, are simultaneously applied to the DUT and reference resistor, respectively.  The signal $\nu_b$ at the so-called ``bridge point'' is then given by
\begin{equation}\label{eq1}
\nu_b = \frac{Y_{DUT}}{Y_\Sigma}\nu_{DUT} + \frac{Y_{REF}}{Y_\Sigma}\nu_{REF},
\end{equation}
where $Y$ refers to the AC admittance and $Y_\Sigma = Y_{DUT} + Y_{par} + Y_{REF}$ is the total admittance seen from the bridge point.  The ``par" subscript refers to parasitic terms, including the HEMT gate impedance.  When balanced, i.e. $\nu_b = 0$, the amplitude and phase of the DUT impedance are given by
\begin{equation}\label{eq2}
Z_{DUT} = \frac{-\nu_{DUT}}{Y_{REF}\cdot\nu_{REF}},
\end{equation}
independent of any parasitic capacitances.  Output characteristics and gate leakage\cite{paranote} (below 1 nA) of the HEMT are shown in Fig.\ref{fig:hemtfig}.  The bias point for the HEMT is chosen to maximize SNR while keeping the DC power as low as possible to avoid thermal drift and temperature gradients during measurement\cite{darknote}.  We found that the optimal bias point for the HEMT is in the triode regime, which has low gain, but also very low noise and thermal drift.  Fig.\ref{fig:noisefig} shows SNR and DC power dissipation for the HEMT as a function of $V_{gs}$ and $V_{dd}$ applied to the 1k$\Omega$ load resistor $R_{load}$, as well as the output power spectrum for the optimal bias point at room temperature for an RMS input AC excitation of 1.5mV at 17.5 kHz.  The spectrum is flat except for low frequencies (below 1kHz), where $1/f$ noise dominates.  We use a Stanford Research Systems lock-in amplifier (model SR830) to recover the small AC output of the amplifier.  We define the output sensitivity $S = \Delta n\sqrt{t_{meas}}$, where $\Delta n$ is the RMS noise in the measured data points sampled with the lock-in, and $t_{meas}$ is the lock-in measurement time\cite{lockinnote} per data point.  At each temperature, we perform a HEMT bias optimization prior to the capacitance measurement.  The optimal bias points with corresponding output sensitivity $S$ for a range of temperatures are given in Table~\ref{tab:tab1}.  The sensitivity improves only slightly with temperature, as the bridge performance is limited by  $1/f$ noise from the HEMT.
\section{Experimental Setup and Measurements}

To demonstrate the performance of our bridge, we compare it against the popular Andeen-Hagerling model 2700A precision capacitance and loss measurement bridge,
the highest resolution CV meter commercially available.  The performance of the commercial bridge was first quantified by measuring a standard shielded 9.3pF capacitor\cite{capnote}.  As expected, the measured sensitivity is a strong function of the test signal ($\sim$200 and $\sim$80aF/$\sqrt{Hz}$ for 100mV and 250mV rms test signals, respectively, in good agreement with the manufacturer specification).  Based on these measurements, more than 10 hours of averaging per data point would be required to obtain 1 attofarad resolution for a 100mV test signal, which still has a peak-to-peak amplitude of more than 10$k_{B}T$ at room temperature.  However, we can avoid long measurement averaging times for the comparison between our bridge and the commercial bridge by measuring the capacitance of graphene devices with strongly-coupled top gates, as the quantum capacitance of such structures can easily be resolved by the commercial bridge.

\begin{figure}
\includegraphics[width=8.5cm]{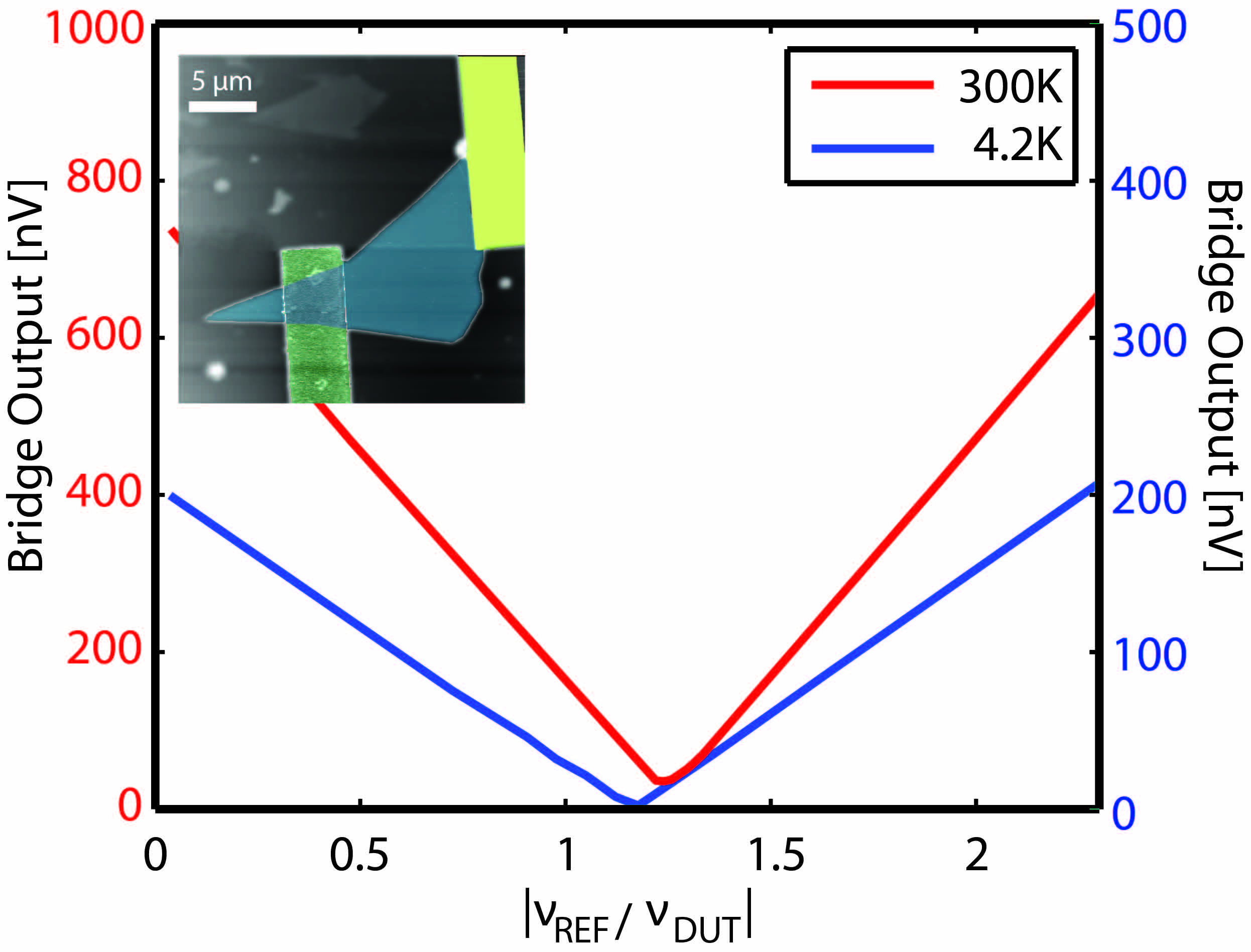}
\caption{\label{fig:zerofig} Integrated bridge output as a function of $|\nu_{REF}/\nu_{DUT}|$ for $T$=300K (left axis), with $\nu_{DUT}$=8mV and phase difference between graphene DUT and REF signals $\Delta\Phi$=144$^\circ$; and for $T$=4.2K (right axis), with $\nu_{DUT}$=100$\mu$V and $\Delta\Phi$=155$^\circ$.  The bridge output minimum and phase difference $\Delta\Phi$ are shifted slightly between the two curves due to changes in the REF and DUT impedances with temperature. The inset shows a false-color AFM image of the measured graphene device.  The graphene is colored blue, the source/drain contact is gold, and the top-gate is colored green.}
\end{figure}

The graphene in our device was deposited on a SiO$_2$/Si chip via mechanical exfoliation, and confirmed to be single-layer via optical contrast and confocal Raman spectroscopy.  100nm of PMMA was spin-coated, and source/drain leads to a selected graphene sheet were defined with e-beam lithography.  Following Ti/Au deposition (5/40 nm, respectively) and liftoff in acetone, the entire chip surface was coated via e-beam evaporation with nominally 1.5 nm of Al, which then oxidized almost immediately upon exposure to air.  A top-gate electrode was patterned on top of the graphene by e-beam lithography, followed by 40nm of e-beam evaporation of Al and liftoff in acetone\cite{geim2010,Ensslin2010}. An AFM image of the device is shown in the inset of Fig.~\ref{fig:zerofig}.  The graphene device chip was finally wirebonded to the bridge chip, and both chips were mounted on a copper support for thermal anchoring.

Measurements were carried out in a Lakeshore/Desert Cryogenics variable temperature probe station.  The HEMT gate impedance $Z_{gate}$ was characterized by measuring the difference in HEMT response between exciting the HEMT gate directly through its small bond pad and exciting the gate through the reference impedance.  The signal attenuation $G$ is then related to the impedances by the equation $G = \frac{Z_{gate}}{Z_{gate} + Z_{REF}}$.  These complex impedance values are used to calculate the relative phase between $\nu_{REF}$ and $\nu_{DUT}$ required to for balancing the bridge.  Bridge balance curves are shown in Fig.\ref{fig:zerofig}.  The change in capacitance of the graphene as the top-gate DC bias is swept (also via Yokogawa 7651) is then calculated from the change in bridge signal $\nu_{b}$ using \eqref{eq1}.

Capacitance from bond pads, wirebonds, and probe tips contribute to the parasitic capacitance $C_{par}$, which is in parallel with $C_{DUT}$ and sets the baseline of the measurement.  The parasitic capacitance remains constant throughout the measurement and is often larger than the change in device capacitance caused by the modulation of carrier density inside the device as a function of gate bias.  We define the measurement sensitivity\cite{snrnote} $S_{meas}\equiv\left|\frac{\partial\nu_{b}}{\partial Y_{DUT}}\right|$, so that we have:
 \begin{equation}\label{eqsens}
\delta C_{DUT} \sim \delta Y_{DUT} = \frac{1}{S_{meas}}\delta\nu_b.
\end{equation}
Evaluating the derivative in \eqref{eqsens} using \eqref{eq1}, we find the expression for the sensitivity:
\begin{equation}\label{eqsens}
S_{meas} = \left|\frac{(\nu_{DUT}-\nu_{REF})Y_{REF} + \nu_{DUT}Y_{par}}{(Y_{DUT}+Y_{REF}+Y_{par})^2}\right|
\end{equation}
Thus, the measurement sensitivity is maximized by increasing $\nu_{DUT}$.  However, in practice, we limit this excitation to $\nu_{DUT}\sim k_{B}T$ to prevent thermal drift and heating, and to avoid blurring of density-of-states features by our excitation.

\section{Results and Discussion}

\begin{figure}
\includegraphics[width=8.5cm]{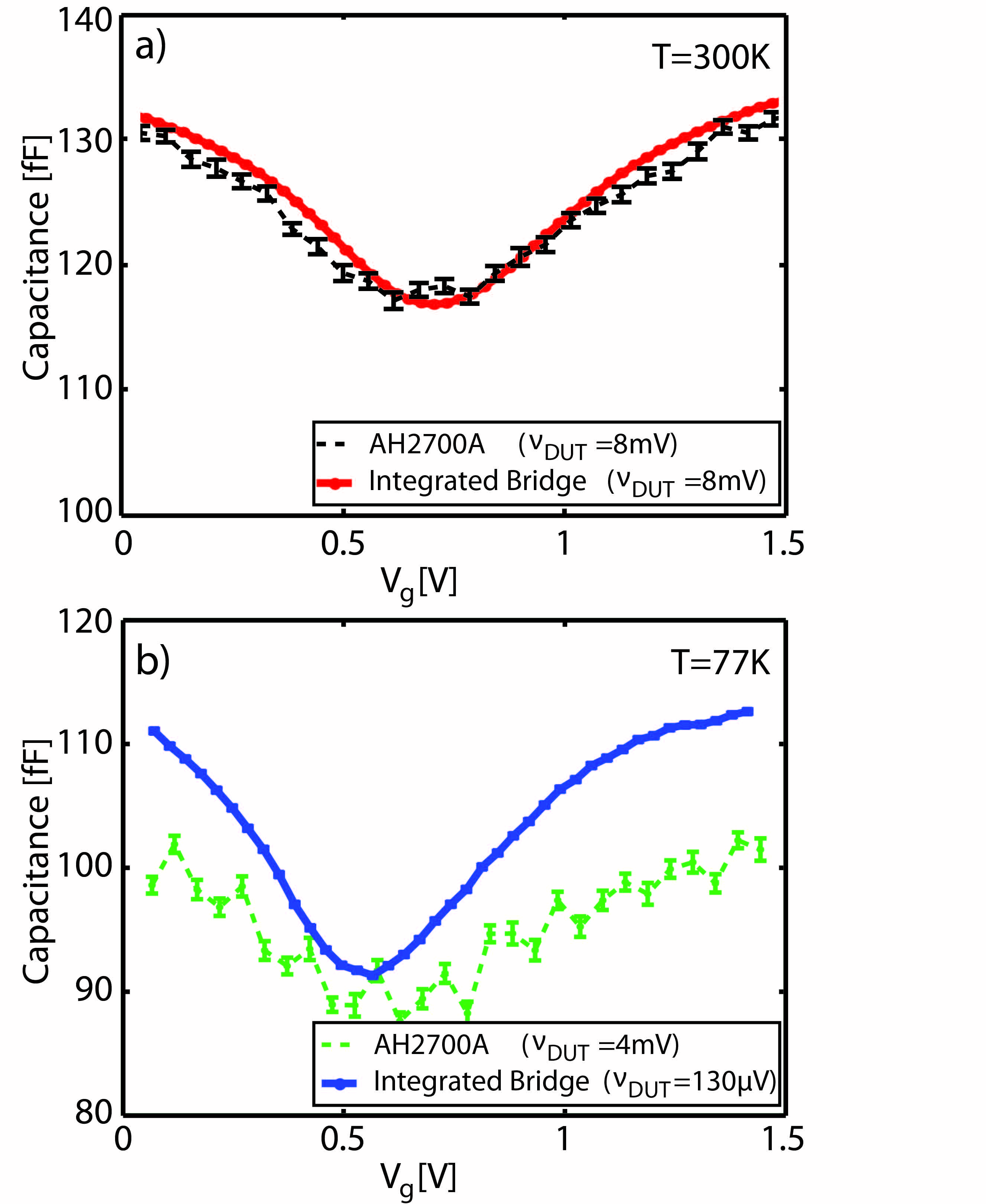}
\caption{\label{fig:graphenecap} CV curves for top-gated graphene device for both the AH2700A and the integrated bridge.  Acquisition time for all measurements was held constant ($\sim$30s per point) to enable direct comparisons between measurements\cite{mnote}.  a) At room temperature using an 8mV DUT excitation, noise in the integrated bridge measurement is $\sim$11aF, while the noise for the commercial bridge is $\sim$675aF.  b)At 77K, the smallest possible DUT excitation of 4mV was used for the commercial bridge measurement, and the Dirac point is poorly resolved due to the high noise level$\sim$5000aF.  For the integrated bridge measurement, the Dirac point is cleanly resolved while using a small DUT excitation of 130$\mu$V, with a noise level$\sim$53aF.}
\end{figure}

2D Graphene has no energy gap between the conduction and valence bands.  Instead, these bands meet at a point, termed the Dirac point, about which the energy-momentum dispersion is linear. The density of states and associated quantum capacitance $C_{Q}$ therefore vary linearly with energy near the Dirac point.  As the top-gate bias voltage is swept, the Fermi level scans the graphene energy spectrum, passing through the Dirac point, where the capacitance reaches a minimum\cite{minnote}.  For top-gate voltages that tune the Fermi level far in energy from the Dirac point, the density of states and associated $C_{Q}$ are large, so that $C_{ox}$ will dominate.

Fig.~\ref{fig:graphenecap}a,b show graphene capacitance measurements for both our integrated bridge and the AH2700A at room temperature and 77K, respectively.  The capacitance is symmetric about the Dirac point, which occurs at a top gate voltage $V_{g}\sim$1V. At room temperature, with an AC excitation of 8mV for $\nu_{DUT}$, the overall capacitance curves match for the two measurements\cite{mnote}, though our integrated bridge is significantly less noisy ($\delta C$ for integrated bridge is $\sim$11aF; for AH2700A, $\delta C\sim$675aF).  At 77K, the minimum allowable excitation amplitude of 4mV was used for $\nu_{DUT}$ for the AH2700A measurements (peak-to-peak amplitude$\sim 2k_{B}T$), while a substantially lower excitation of 130$\mu$V was used for $\nu_{DUT}$ for the integrated bridge measurements (peak-to-peak amplitude$\sim k_{B}T/18$).  The capacitance curve is again cleanly resolved for the integrated bridge measurements ($\delta C\sim$53aF), with improvement in the sharpness of the capacitance around the Dirac point due to lower temperature and excitation (depth of Dirac point capacitance dip $\Delta C$ increased from $\sim$15fF to $\sim$21fF), whereas these features are obscured by the excessive noise level for the AH2700A measurements ($\delta C\sim$5000aF).

\section{Conclusions}
We have demonstrated a reliable method for integrated high-resolution quantum capacitance measurements over a wide temperature range using an integrated bridge circuit directly wirebonded to the DUT.  The performance of our bridge was tested against the commercially available AH2700A capacitance bridge by measuring the capacitance of a top-gated graphene device.  We observed significant resolution improvements over the commercial tool, enabling the fast measurement of quantum capacitance for nanostructures down to cryogenic temperatures, and achieving 10s of attofarad resolution per root hertz at room temperature (equivalently, less than 1 e$^\mathrm{-}$ per root hertz on the DUT) while limiting the excitation amplitude to below $k_{B}T$.

\begin{acknowledgements}
We thank Ray Ashoori, Stuart Tessmer, and Gary Steele for their pioneering work in this field, and for many helpful discussions.  We also thank James Harris and Shahal Ilani for measurement advice, YiChing Pao and Diana Fong for help with substrate preparation and wirebonding, and Jeff Bokor and Patrick Bennett for discussions and use of a 2$^{nd}$ AH2700A.  This work was supported by the FENA Focus Center, one of the six research centers funded under the Focus Center Research Program (FCRP), a Semiconductor Research Corporation subsidiary, as well as by the Air Force Office of Scientific Research (contract FA9550-08-1-0427).
\end{acknowledgements}

\nocite{*}

\begin{thebibliography}{16}%
\makeatletter
\providecommand \@ifxundefined [1]{%
 \@ifx{#1\undefined}
}%
\providecommand \@ifnum [1]{%
 \ifnum #1\expandafter \@firstoftwo
 \else \expandafter \@secondoftwo
 \fi
}%
\providecommand \@ifx [1]{%
 \ifx #1\expandafter \@firstoftwo
 \else \expandafter \@secondoftwo
 \fi
}%
\providecommand \natexlab [1]{#1}%
\providecommand \enquote  [1]{``#1''}%
\providecommand \bibnamefont  [1]{#1}%
\providecommand \bibfnamefont [1]{#1}%
\providecommand \citenamefont [1]{#1}%
\providecommand \href@noop [0]{\@secondoftwo}%
\providecommand \href [0]{\begingroup \@sanitize@url \@href}%
\providecommand \@href[1]{\@@startlink{#1}\@@href}%
\providecommand \@@href[1]{\endgroup#1\@@endlink}%
\providecommand \@sanitize@url [0]{\catcode `\\12\catcode `\$12\catcode
  `\&12\catcode `\#12\catcode `\^12\catcode `\_12\catcode `\%12\relax}%
\providecommand \@@startlink[1]{}%
\providecommand \@@endlink[0]{}%
\providecommand \url  [0]{\begingroup\@sanitize@url \@url }%
\providecommand \@url [1]{\endgroup\@href {#1}{\urlprefix }}%
\providecommand \urlprefix  [0]{URL }%
\providecommand \Eprint [0]{\href }%
\@ifxundefined \urlstyle {%
  \providecommand \doi  [0]{\begingroup \@sanitize@url \@doi}%
  \providecommand \@doi [1]{\endgroup \@@startlink {\doibase
  #1}doi:\discretionary {}{}{}#1\@@endlink }%
}{%
  \providecommand \doi  [0]{doi:\discretionary{}{}{}\begingroup
  \urlstyle{rm}\Url }%
}%
\providecommand \doibase [0]{http://dx.doi.org/}%
\providecommand \Doi [0]{\begingroup \@sanitize@url \@Doi }%
\providecommand \@Doi  [1]{\endgroup\@@startlink{\doibase#1}\@@Doi}%
\providecommand \@@Doi [1]{#1\@@endlink}%
\providecommand \selectlanguage [0]{\@gobble}%
\providecommand \bibinfo  [0]{\@secondoftwo}%
\providecommand \bibfield  [0]{\@secondoftwo}%
\providecommand \translation [1]{[#1]}%
\providecommand \BibitemOpen [0]{}%
\providecommand \bibitemStop [0]{}%
\providecommand \bibitemNoStop [0]{.\EOS\space}%
\providecommand \EOS [0]{\spacefactor3000\relax}%
\providecommand \BibitemShut  [1]{\csname bibitem#1\endcsname}%
\bibitem [{itr(2009)}]{itrs2009}%
  \BibitemOpen
  \href@noop {} {\enquote {\bibinfo {title} {International technology roadmap
  for semiconductors},}\ }\bibinfo {type} {Tech. Rep.}\ (\bibinfo
  {institution} {ITRS},\ \bibinfo {year} {2009})\BibitemShut {NoStop}%
\bibitem [{\citenamefont {Datta}(2007)}]{datta2007}%
  \BibitemOpen
  \bibfield  {author} {\bibinfo {author} {\bibfnamefont {S.}~\bibnamefont
  {Datta}},\ }\href@noop {} {\emph {\bibinfo {title} {Quantum Transport: Atom
  to Transistor}}}\ (\bibinfo  {publisher} {Cambridge University Press},\
  \bibinfo {year} {2007})\BibitemShut {NoStop}%
\bibitem [{\citenamefont {Lee}(1982)}]{Lee1982}%
  \BibitemOpen
  \bibfield  {author} {\bibinfo {author} {\bibfnamefont {P.~A.}\ \bibnamefont
  {Lee}},\ }\bibfield  {title} {\enquote {\bibinfo {title} {Density of states
  and screening near the mobility edge},}\ }\href@noop {} {\bibfield  {journal}
  {\bibinfo  {journal} {Phys.\ Rev.\ B},\ }\textbf {\bibinfo {volume} {26}},\
  \bibinfo {pages} {5882} (\bibinfo {year} {1982})}\BibitemShut {NoStop}%
  \bibitem [{\citenamefont {lur}(1988)}]{luryi}%
  \BibitemOpen
  \bibfield  {author} {\bibinfo {author} {\bibfnamefont {S.}\ \bibnamefont
  {Luryi}},\ }\bibfield  {title} {\enquote {\bibinfo {title} {Quantum capacitance devices},}\ }\href@noop {} {\bibfield  {journal}
  {\bibinfo  {journal} {Appl.\ Phys.\ Lett.},\ }\textbf {\bibinfo {volume} {52}},\
  \bibinfo {pages} {501} (\bibinfo {year} {1988})}\BibitemShut {NoStop}%
\bibitem [{\citenamefont {Ilani}\ \emph {et~al.}(2006)\citenamefont {Ilani},
  \citenamefont {Donev}, \citenamefont {Kindermann},\ and\ \citenamefont
  {McEuen}}]{Ilani2006}%
  \BibitemOpen
  \bibfield  {author} {\bibinfo {author} {\bibfnamefont {S.}~\bibnamefont
  {Ilani}}, \bibinfo {author} {\bibfnamefont {L.~A.~K.}\ \bibnamefont {Donev}},
  \bibinfo {author} {\bibfnamefont {M.}~\bibnamefont {Kindermann}}, \ and\
  \bibinfo {author} {\bibfnamefont {P.~L.}\ \bibnamefont {McEuen}},\ }\bibfield
   {title} {\enquote {\bibinfo {title} {Measurement of the quantum capacitance
  of interacting electrons in carbon nanotubes},}\ }\href@noop {} {\bibfield
  {journal} {\bibinfo  {journal} {Nat Phys},\ }\textbf {\bibinfo {volume}
  {2}},\ \bibinfo {pages} {687} (\bibinfo {year} {2006})}\BibitemShut {NoStop}%
\bibitem [{\citenamefont {Steele}(2006)}]{steele}%
  \BibitemOpen
  \bibfield  {author} {\bibinfo {author} {\bibfnamefont {G.}~\bibnamefont
  {Steele}},\ }\emph {\bibinfo {title} {Imaging Transport Resonances in the
  Quantum Hall Effect}},\ \href@noop {} {\bibinfo {type} {{PhD}
  dissertation}},\ \bibinfo  {school} {Massachusetts Institute of Technology},
  \bibinfo {address} {Department of Physics} (\bibinfo {year}
  {2006})\BibitemShut {NoStop}%
\bibitem [{\citenamefont {Urazhdin}\ \emph {et~al.}(2002)\citenamefont
  {Urazhdin}, \citenamefont {Tessmer},\ and\ \citenamefont
  {Ashoori}}]{urazhdin2002}%
  \BibitemOpen
  \bibfield  {author} {\bibinfo {author} {\bibfnamefont {S.}~\bibnamefont
  {Urazhdin}}, \bibinfo {author} {\bibfnamefont {S.~H.}\ \bibnamefont
  {Tessmer}}, \ and\ \bibinfo {author} {\bibfnamefont {R.~C.}\ \bibnamefont
  {Ashoori}},\ }\bibfield  {title} {\enquote {\bibinfo {title} {A simple
  low-dissipation amplifier for cryogenic scanning tunneling microscopy},}\
  }\href@noop {} {\bibfield  {journal} {\bibinfo  {journal} {Rev. Sci.
  Instrum.},\ }\textbf {\bibinfo {volume} {73}},\ \bibinfo {pages} {310}
  (\bibinfo {year} {2002})}\BibitemShut {NoStop}%
\bibitem [{res()}]{resnote}%
  \BibitemOpen
  \href@noop {} {}\bibinfo {note} {Several other commercially available HEMTs,
  including the Agilent ATF 33143 and 34143, were found to be unsuitable in our
  measurements due to high-frequency (MHz) transport resonances.}\BibitemShut
  {Stop}%
\bibitem [{par()}]{paranote}%
  \BibitemOpen
  \href@noop {} {}\bibinfo {note} {Measured via an Agilent B1500A parameter
  analyzer and Keithley 2612.}\BibitemShut {Stop}%
\bibitem [{dar()}]{darknote}%
  \BibitemOpen
  \href@noop {} {}\bibinfo {note} {All measurements were performed in the dark
  to prevent optical excitation of the exposed 2DEG in the HEMT.  After cooling, HEMTs were temporarily exposed to light, to enable navigating probes to pads, so persistent photoconductivity may affect HEMT characteristics at 77K and 4K.}\BibitemShut
  {Stop}%
\bibitem [{loc()}]{lockinnote}%
  \BibitemOpen
  \href@noop {} {}\bibinfo {note} {Measurement time $t_{meas}$ is proportional
  to the lock-in time constant and filter slope (SRS830 manual).}\BibitemShut
  {Stop}%
\bibitem [{cap()}]{capnote}%
  \BibitemOpen
  \href@noop {} {}\bibinfo {note} {The standard capacitor was measured using
  two different AH2700A units in two separate locations.}\BibitemShut {Stop}%
\bibitem [{\citenamefont {Ponomarenka}\ \emph {et~al.}(2010)\citenamefont
  {Ponomarenka}, \citenamefont {Yang}, \citenamefont {Gorbachev}, \citenamefont
  {Blake}, \citenamefont {Katsnelson}, \citenamefont {Novoselov},\ and\
  \citenamefont {Geim}}]{geim2010}%
  \BibitemOpen
  \bibfield  {author} {\bibinfo {author} {\bibfnamefont {L.~A.}\ \bibnamefont
  {Ponomarenka}}, \bibinfo {author} {\bibfnamefont {R.}~\bibnamefont {Yang}},
  \bibinfo {author} {\bibfnamefont {R.~V.}\ \bibnamefont {Gorbachev}}, \bibinfo
  {author} {\bibfnamefont {P.}~\bibnamefont {Blake}}, \bibinfo {author}
  {\bibfnamefont {M.~I.}\ \bibnamefont {Katsnelson}}, \bibinfo {author}
  {\bibfnamefont {K.~S.}\ \bibnamefont {Novoselov}}, \ and\ \bibinfo {author}
  {\bibfnamefont {A.~K.}\ \bibnamefont {Geim}},\ }\href@noop {} {\enquote
  {\bibinfo {title} {Density of states and zero landau level probed through
  capacitance of graphene},}\ }\bibinfo {howpublished} {e-print
  arXiv:1005.4793v1} (\bibinfo {year} {2010})\BibitemShut {NoStop}%
\bibitem [{\citenamefont {Droscher}\ \emph {et~al.}(2010)\citenamefont
  {Droscher}, \citenamefont {Roulleau}, \citenamefont {Molitor}, \citenamefont
  {Studerus}, \citenamefont {Stampfer}, \citenamefont {Ihn},\ and\
  \citenamefont {Ensslin}}]{Ensslin2010}%
  \BibitemOpen
  \bibfield  {author} {\bibinfo {author} {\bibfnamefont {S.}~\bibnamefont
  {Droscher}}, \bibinfo {author} {\bibfnamefont {P.}~\bibnamefont {Roulleau}},
  \bibinfo {author} {\bibfnamefont {F.}~\bibnamefont {Molitor}}, \bibinfo
  {author} {\bibfnamefont {P.}~\bibnamefont {Studerus}}, \bibinfo {author}
  {\bibfnamefont {C.}~\bibnamefont {Stampfer}}, \bibinfo {author}
  {\bibfnamefont {T.}~\bibnamefont {Ihn}}, \ and\ \bibinfo {author}
  {\bibfnamefont {K.}~\bibnamefont {Ensslin}},\ }\bibfield  {title} {\enquote
  {\bibinfo {title} {Quantum capacitance and density of states of graphene},}\
  }\href@noop {} {\bibfield  {journal} {\bibinfo  {journal} {Appl.\ Phys.\
  Lett.},\ }\textbf {\bibinfo {volume} {96}},\ \bibinfo {pages} {152104}
  (\bibinfo {year} {2010})}\BibitemShut {NoStop}%
\bibitem [{snr()}]{snrnote}%
  \BibitemOpen
  \href@noop {} {}\bibinfo {note} {The signal-to-noise ratio is related to the measurement sensitivity by $SNR = S_{meas}\frac{Y_{DUT}}{\nu_b}$.}\BibitemShut {Stop}%
\bibitem [{mno()}]{mnote}%
  \BibitemOpen
  \href@noop {} {}\bibinfo {note} {The acquisition time for each data point
  ($\sim$30s) is approximately equal for both the integrated bridge and the
  AH2700A measurements, allowing for direct comparison between curves. Note
  that this per-point acquisition time contains settling times as well as wait
  times associated with polling the GPIB bus.}\BibitemShut {Stop}%
\bibitem [{min()}]{minnote}%
  \BibitemOpen
  \href@noop {} {}\bibinfo {note} {The minimum capacitance is limited by the
  temperature, disorder, and parasitic capacitance\cite{geim2010}.}\BibitemShut
  {Stop}%

\end{thebibliography}
\providecommand{\noopsort}[1]{}\providecommand{\singleletter}[1]{#1}%

\end{document}